\documentclass[11pt]{llncs}

\usepackage{epsf}
\usepackage{amsbsy}
\usepackage{epsfig}
\usepackage{amssymb}
\usepackage{mathrsfs}
\usepackage{latexsym}
\usepackage{enumerate}
\usepackage{amsmath}
\usepackage{amsfonts}

\newcommand{\remove}[1]{}
\newcounter{pclaim} 

\begin{document}

\title{\LARGE\bf Faster Gossiping in Bidirectional Radio Networks with Large Labels}

\author{
\sc{Shailesh Vaya} \institute{Department of Computer Science and Engineering, \\ Indian Institute of Technology Patna}
\email{\\shailesh.vaya@gmail.com}.
}


\maketitle


\abstract
{
  We consider unknown ad-hoc radio networks, when the underlying network is bidirectional and nodes can have polynomially large labels. For this model, we present a deterministic protocol for gossiping which takes $O(n \lg^2 n \lg \lg n)$ rounds. This improves upon the previous best result for deterministic gossiping for this model by [Gasienec, Potapov, Pagourtizis, Deterministic Gossiping in Radio Networks with Large labels, ESA (2002)], who present a protocol of round complexity $O(n \lg^3 n \lg \lg n)$ for this problem. This resolves open problem posed in [Gasienec,  Efficient gossiping in radio networks, SIROCCO (2009)], who cite bridging gap between lower and upper bounds for this problem as an important objective. We emphasize that a salient feature of our protocol is its simplicity, especially with respect to the previous best known protocol for this problem.
}
\newline



\setcounter{page}{2}

\section{Introduction}
\label{sec:introduction}
  Mobile ad-hoc radio networks play an important role in a wide range of fields, ranging from agriculture and automobiles to hazardous environments and defense. Often enough, radio networks and their connectivity are not planned ahead precisely and there is often no centralized system to coordinate the deployed radio stations. Thus, typically radio networks form a classical distributed setting in which the radio nodes have to rely solely on the communication received by it and its own label (possibly knowledge of some other parameters) to determine its action at any time step. Complex communication activities that the deployed radio networks may undertake relies on fundamental communication primitives like broadcasting and gossiping.

  In the {\it broadcasting problem}, a message from a distinguished source node is to be communicated to all other nodes of the network. While in the {\it gossiping} problem, each node of the network may possess a different message which is to be communicated to rest of the network. Needless to say, while broadcasting can be achieved on any directed network, gossiping can be completed only on strongly connected networks. For both the problems, the most important consideration is the time from initiation of the task to completion of the task. However, in an asynchronous system, real clock time with respect to an external entity is an improper measure. In this work, we assume a synchronous model with no faulty nodes, when nodes can communicate messages of arbitrary sizes. Computation and communication for this model is organized in rounds. In every round, each node can either act as a transmitter or a receiver, but not both. This is an important distinguishing feature of radio networks. Whether a node is actually able to receive a message or the message sent by it is received by a neighboring node depends on another salient property of radio networks: {\em If two in-neighbors of a node transmit any message in the same round, then a collision occurs and the receiving node receives nothing. In particular, the receiving node cannot distinguish it from the case when all of its in-neighbors were silent}. It is this features of radio networks that make even a simple task of broadcasting non-trivial.

  Finally, the (in)availability of connectivity knowledge is an important consideration under which radio broadcasting and gossiping problem have been studied. It is typically assumed that nodes do not have any connectivity information about the network, but perhaps knowledge of $n$ - the number of nodes in the network, $\Delta$ - the maximum in-degree of a node in the network, or $D$ - the diameter of the network. However, radio broadcasting and gossiping problem have also been studied under the assumption that nodes in the network do not have any apriori knowledge of any of these parameters, or even $n$, which is a valid assumption for many settings.

  There are several metrics to measure the complexity of broadcasting and gossiping protocols, like round complexity (minimum number of rounds needed to complete broadcasting), message complexity (minimum number of messages needed to be sent to complete broadcasting), etc. These metrics are defined in terms of the number of nodes in the network $n$ and the radius of the network $D$. The most important and prevalent complexity measure under which the radio broadcasting and gossiping problem have been studied is the {\it round complexity}. This will be the criterion by which we will calibrate broadcasting and gossiping protocols in this work also.

  Early works for broadcasting and gossiping also made the strong assumption that the labels of nodes of the network belong to a small range $[1,\dots,O(n)]$. This is restrictive and not very practical. This restriction on label values allows the parties to engage in a simple round-robin protocol, where they all transmit their ID's one by one in a separate round and eventually learn about their in-neighborhoods etc.. Allowing nodes to have large labels causes the round-robin phase in broadcasting/gossiping protocols to cost $O(n^c)$ rounds, making them very time expensive. This brings into consideration new challenges in protocol design. Peleg initiated discussion on this issue in \cite{P} and it has been considered important that protocols work for the case when nodes can have polynomial sized labels.

  In \cite{GPP02}, authors consider the problem of deterministic gossiping in unknown radio networks, when underlying network is strongly connected networks or bi-connected, and nodes can have polynomially large labels. For this model, a deterministic gossiping protocol of $\tilde{O}(n^{5/3})$ round complexity is presented for strongly connected networks. For bidirectional networks, a gossiping protocol is presented which takes $O(n \lg^4 n)$ rounds. This protocol proceeds in phases, in which the nodes learn about the neighborhood before executing a depth first search on the graph that takes $O(n)$ rounds. The phase for learning the neighborhood invokes leader election protocol on the network $O(\lg n)$ times, which in turns invokes deterministic broadcasting protocol $O(\lg n)$ times. Using the current best protocol for deterministic radio broadcasting, in \cite{M10}, which takes $O(n \lg n \lg \lg n)$ rounds and works for the case of large labels, the round complexity of gossiping protocol for bidirectional networks in \cite{GPP02} takes $O(n \lg^3 n \lg \lg n)$ rounds. The significance of reducing the gap between upper and lower bounds for this problem has been further emphasized in the survey article by Gasienec, \cite{G09}, who pose this as an important goal.

  In this work, we present a deterministic protocol for radio gossiping in unknown bidirectional networks which takes $O(n \lg^2 n \lg \lg n)$ rounds. The shaving off $O(\lg n)$ factor, from the round complexity of previous best protocol given in \cite{GPP02}, is achieved by reducing the number of invocations of the leader election algorithm to only once.

\subsection{Previous Works}
\label{sec:related-wonk}
  The study of broadcasting has received considerably more attention then gossiping. For directed graphs, a series of works improved the trivial upper bound of $O(n^2)$ for broadcasting (for small labels), to $O(n^{11/16})$ by \cite{CGGPR02}, to $\tilde{O}(n^{5/3})$ by tDe Marco and Pelc, \cite{MP01}, $O(n^{3/2})$ by \cite{CGOR00}, $O(n \lg^2 n)$ by \cite{CGR02}, $O(n \lg^2 D)$ by \cite{CR} and most recently $O(n \lg n \lg \lg n)$ by \cite{M10}. \cite{BP} showed that for any deterministic algorithm $\mathcal{A}$ for broadcasting in ad-hoc radio networks, there are networks on which $\mathcal{A}$ requires $\Omega(n \lg n)$ rounds.

  The first sub-quadratic deterministic algorithm for the gossiping problem in ad-hoc radio networks was $\tilde{O}(n^{3/2})$ time algorithm proposed in \cite{CGR02}. Subsequently, \cite{Xu}, improved this bound by a poly-logarithmic factor obtaining $O(n^{3/2})$ bound. For small diameter $D$, the gossiping time was later improved by Gasieniec and Lingas \cite{GL02} to $\tilde({O}(nD^{1/2})$. These algorithms assume that the node labels are linear in $n$. Clementi, Monti, and Silvestri \cite{CMS} presented a $\tilde{O}(D {\Delta}^2)$-time deterministic algorithm, which was improved to $\tilde{O}(D \Delta^{3/2})$ by Gasieniec and Lingas \cite{GL02}, for polynomially large labels. \cite{GRX04} improved the result on deterministic gossiping from $\tilde{O}(n^{5/3})$ in \cite{GPP02} to $\tilde{O}(n^{4/3})$ in \cite{GRX04}. An excellent survey of results on deterministic gossiping problem is presented by Gasieniec in \cite{G09}. For bounded size messages deterministic gossiping was studied in \cite{CGL02} and for unit size messages gossiping protocols were given in \cite{GP} and \cite{MX}. \cite{CGR02} proposed an $O(n \lg^4 n)$ time randomized gossiping algorithm, which was improved to $O(n \lg^3 n)$ rounds by\cite{LP} and $O(n \lg^2 n)$ rounds by \cite{CR}.

  \cite{CGGPR02}, \cite{OCW03}, \cite{UCW07} study the problem of acknowledged broadcasting and gossiping when nodes do not know any estimate on the number of nodes in the network $n$. \cite{FP08} improve their results and present an $O(n \lg n \lg \lg n)$-time algorithm for deterministic broadcasting, for the case of small labels.

  In the centralized setting, the topology of the radio network is known to all the nodes of the network. For this setting, a deterministic broadcasting protocol of $O(D \lg^2 n)$ was given in \cite{CW}, for networks of radius $D$. For the centralized setting, a lower bound of $\Omega(D + \lg^2 n)$ rounds was proved in \cite{ANLP}. This was shown to be tight in a recent work \cite{KP5}, where a matching upper bound of $\Omega(D + \lg^2 n)$ rounds was proved, after a series of improvements from $O(D \lg^2 n)$ rounds in \cite{CW}), $O(D \lg n + \lg^2 n)$ rounds in \cite{KP4}, $O(D + \lg^4)$ rounds in \cite{EK} and $O(D + \lg^5 n)$ rounds in \cite{GM}.

\subsection{Organization of the paper}
  In Section \ref{sec:model-definitions}, we present the model, some relevant tools and terminologies. In Section \ref{sec:discover}, we review some primitives for bi-directional networks introduced in \cite{KP04}. We adapt them and employ in our protocols for bidirectional networks. Finally, in Section \ref{sec:fastgpbi}, we present our new deterministic algorithm for gossiping in bidirectional networks.

\section{Description of the model and relevant tools}
\label{sec:model-definitions}

\begin{definition} (\cite{BGI},\cite{KP04})
  A broadcast protocol $\pi$ for a radio network is a synchronous multi-processor protocol which proceeds in rounds. Following are the salient features of the broadcasting process for undirected radio networks:
\label{definition}
\begin{enumerate}
\item
   All nodes execute identical copies of the same protocol $\pi$.
\item
   In each round, every node either acts as a transmitter or as a receiver (or is inactive).
\item
   A node receives a message in a specific round if and only if it acts as a receiver and exactly one of its neighbors transmits in that round. Otherwise, it receives $\phi$.  We assume that the messages are authenticated, that is, when a node receives a message it gets to know the label of the transmitting node.
\item
   The action of a node in a specific round is determined by
   \begin{enumerate}
   \item
      Initial input, which typically contains own label, and sometimes also the number of nodes in the networks $n$.
   \item
      Messages received by the node in previous rounds.
   \end{enumerate}
\item
   Broadcast is completed in $r$ rounds if all the nodes receive the source message in one of the rounds $0,1,\dots,r-1$.
\end{enumerate}
\end{definition}

\subsection{Combinatorial tools}
\label{sec:tools}
  We briefly review a few different combinatorial tools used in this work.

\begin{definition} (Selective family)
\label{selectivefamily}
  A ($k,n$)-selective family $\mathcal{F}$ consists of a set of subsets of a set $S = [1,\dots,n]$, such that for every subset $s_k$ of $S$ of size at most $k$, there exists at least one subset $s_f \in \mathcal{F}$ for which $|s_f \bigcap s_k| = 1$.
\end{definition}

  The number of subsets $\mathcal{F}$ in a ($k,n$)-selective family is denoted by $SF(k,n)$.

\begin{theorem}
\label{theo:selectivefamily}
  There exists a ($k,n$)-selective family $\mathcal{F}$ of size $f \cdot k \cdot \lg \frac{n}{k}$, for some constant $f$.
\end{theorem}

  The following result was presented in \cite{M10} for deterministic broadcasting in directed networks and will be used by us:
\begin{theorem}
\label{fastbroadcast}
  There exists a deterministic broadcasting algorithm that works in $O(n \lg n \lg \lg n)$ rounds on all directed graphs. When $n$ nodes of the network can have labels in range $[1, \dots, n^c]$, the algorithm is referred to by $RB(n,n^c)$ and the number of rounds taken by it, by $NB(n,n^c)$.
\end{theorem}

\section{Faster gossiping in bi-directional networks with large labels}
\label{sec:fastgpbi}
  First, we present a few primitives for bidirectional networks which will be used in our main protocol. These primitives have been adapted from \cite{KP04} and their correctness follows from their.

\subsection{Primitives for bidirectional networks}
\label{sec:discover}
  In this section, we present a sub-protocol which will be used in composing some other protocols. This protocol has been obtained by adapting the Binary-Selection-Broadcast() procedures presented in \cite{KP04}. 

 The protocol described in SubSubsection \ref{echo} is used to estimate the size of set $A$: Determine if the number of nodes in subset $A$ connected to a node $s$ is $0$, $1$ or more. This is achieved with help of a helper node whose label is $h$. This is the similar to the Echo() procedure given in \cite{KP04}. SubSubsection \ref{binaryselectionbroadcast}, presents a binary selection procedure for a node to execute along with its neighboring nodes to select an undiscovered neighboring node from them.

\subsubsection{Estimating the size of a neighborhood}
\label{echo}
  Procedure Estimate() is initiated by a node $s$. It works with three parameters: (1) $X$ which is a subset of labels to be excluded from the search (2) $h$ which is the label of a particular node (3) $Y$ a subset of labels within which the search is to be carried out.\\

\underline{Procedure Estimate($h$, $X$, $Y$)}
\begin{enumerate}
\item{Step 0.} Node $s$ transmits a message $h, X, Y$. This message is received by all its neighbors.

\item{Step 1} Every neighbor of $s$ with labels in subset $Y - X - \{h\}$ transmits its label.

\item{Step 2} Every neighbor of $s$ in $Y - X \bigcup \{h\}$ transmits its label.
\begin{enumerate} There are 3 possible effects of executing Estimate($h$, $X$, $Y$) at node $s$.
\item{Case 1} A message is received in Step 1 and no message is received in Step 2. In this case, the node $s$ knows that it has a single neighbor with label in $Y - X -\{h\}$ and learns the label of this unique node.
\item{Case 2} No message is received in Step 1 but a message from $h$ is received in Step 2. In this case, the node $s$ knows that it has no neighbors with labels in $Y - X -\{h\}$.
\item{Case 3} No message is received in either Step. In this case, the node $s$ concludes that $s$ has at least $2$ neighbors in $Y - X -\{h\}$.
\end{enumerate}

\end{enumerate}

\subsubsection{Binary selection of an undiscovered node}
\label{binaryselectionbroadcast}
  The goal of this procedure is for a node $s$, to discover a new node in its undiscovered neighborhood $\overline(X)$, with the help of an (already discovered) assistant node $h$.\\

\underline{Binary-Select($s,h,X$)}\\
\begin{enumerate}
\item{Step $0$:} Node $s$ transmits the label of a neighbor node $h$, which is called helper of $s$ and subset of excluded labels $X$. 

\item{Step $1$:} $s$ first determines if there exists an undiscovered node in its neighborhood by executing Estimate($h$, $X$, $\inf$).
  If it is determined that the entire neighborhood of $s$ has been discovered, then the procedure Binary-Select() terminates, else continue with Step $2$.

\item{Step $2$:}
\begin{enumerate}
\item{(a.)} Initialize $i$, so that $2^i$ is the minimum $i$ for which $2^i \geq n$.
\item{(b.)} Nodes execute Estimate($h$, $X$, $[1,\dots,n]$) initiated by $s$.
\item{(c.)} If a single new neighbor is discovered, then its label is learned by $s$ and the procedure terminates;\\
       Else, if it is discovered that the number of neighbors of $s$ is $\geq 2$, then Step $3$ terminates and continue to Step $3$.\\
       Else, $i = i+1$ and Sub-Steps (b) and (c) of Step $2$ are executed with incremented value of $i$.
\end{enumerate}

  At termination of Step $2$, nodes conclude that at least $2$ undiscovered neighbors of $s$ have labels in the range $[1,\dots,2^i]$ and continue the search in this range.

  Each of the remaining Steps $3, 4, 5, \dots$ consists of the following binary search stage: Nodes execute Procedure Estimate($h$, $X$, $Y$), with range $Y = [a=1,\dots,b=2^{i-1}]$), initiated by $s$. If the answer received from the execution is the label of a single undiscovered neighbor of $s$, then the procedure terminates and the value of this label is returned; If the answer received from the execution is $0$, then in the next Step nodes continue the search on complimentary range $Y = [a=2^{i-1}+1,\dots,b=2^i]$, else the nodes continue the search on range $Y = [a=1,\dots,b=2^{i-2}]$.
\end{enumerate}

\subsection{Gossiping in bi-directional networks with large labels in $O(n \lg^2 n \lg \lg n)$}
\label{sec:protocol}
  The protocol proceeds in four stages: In the First stage, the nodes execute a binary search procedure to elect a node with the maximum ID as a leader. In the Second and Third stages, the leader conducts a broadcast on the network collecting messages from various nodes of the network. Finally, the leader conducts another broadcast of the collected messages, communicating all the collected messages to the rest of the network.\\

\underline{Protocol for radio gossiping on bidirectional networks}\\
\begin{enumerate}
\item{SELECT-A-LEADER:}
  This stage consists of the following steps conducted iteratively for $j = 1, 2, \dots, \lg n^c$. In step $1$, nodes know that a leader exists with label in $[1, \dots, n^c]$. All nodes with labels in $[\lfloor \frac{n^c}{2} \rfloor, \dots, n^c]$ initialize themselves with a message "1". Nodes, then execute protocol $RB(n+1,n^c)$ for radio broadcasting to disperse this message in the entire network. If a "1" is received by all nodes, by round $ $, then all nodes know that leader exists in $[\lfloor \frac{n^c}{2} \rfloor, \dots, n^c]$. Nodes then continue to search for the leader in this smaller range. Else, the nodes search for leader in range $[1, \dots, \lfloor \frac{n^c}{2}]$.

  This step is executed $\lg_2 n^c$ times, after which all nodes and the leader know the ID of the leader. This stage takes a total of $O(\lg n \cdot NB(n+1, n^c))$ rounds.

\item{DESIGNATE-HELPER-FOR-LEADER:}
  The leader elected in the last Stage is the source node which will conduct the rest of the protocol. From, the messages received by it in the previous Stage from various nodes, the source arbitrarily chooses one node and designates it as its helper node. The source, then re-announces this fact to its neighborhood by transmitting it once.

  Alternatively, the leader transmits a message once to its entire neighborhood, who then execute a $(n,n)$-selective family. Leader then chooses a neighboring node and demarcates it as a helper node $h$.

\item{MARK-AND-PASS-TOKEN:}
  This stage is executed inductively. Subset $X$ is used to maintain already marked nodes in the network. $X$ is passed along with the token in the future steps.

\begin{enumerate}
\item{(First Step:)} The subset $X$ is initialized: $X = \{ s \}$. The token is passed to node $h$, which was designated as the 'helper' in the previous Stage.

\item{($i^{th}$ Step:)} Let $r$ be the node which received the token in the previous Step, from some node a $r_h$. Then, update subset $X$: $X = X \bigcup \{ r \}$. $r$ transmits the message $m$ to be broadcasted. Using the helper node $r_h$, node $r$ attempts to discover a new neighbor by execution of Binary-Select($r,h, X$). The result of this procedure can be one the following:\\ 
\begin{enumerate}
\item Let $t$ be a new node discovered by the Binary Selection procedure. Then, the token is passed to the newly discovered node $t$ and the node from which the token is passed i.e., $r$ is identified as a helper to node $t$ and this step is executed again.

\item If no new node is discovered, then the token is returned back (along with subset $X$) to the node $r_h$ from which node $r$ had received the token for the first time.
\end{enumerate}
\end{enumerate}

\end{enumerate}

  The correctness of the above algorithm follows from correctness of broadcasting protocol \cite{M10} which is used to construct the leader election procedure, as in \cite{CGR02}. The rest of the routines use ideas from algorithm presented in previous Section. We have the following result:

\begin{theorem}
  There exists a deterministic protocol that takes $O(n \lg^2 n \lg \lg n)$ rounds to complete gossiping on every bidirectional network of $n$ nodes, when node labels can take polynomially values in $n$.
\end{theorem}



\begin{thebibliography}{02}
{\scriptsize
\bibitem{ANLP}
Noga Alon, Amok Bar-Noy, Nathan Linial and David Peleg.
{\em A lower bound for radio broadcast},
Journal of Computer and System Sciences 43(2): 290-298 (1991).

\bibitem{A}
Baruch Awerbuch.
{\em A new distributed depth-first-search algorithm},
Information Processing Letters 20: 147-150 (1985).

\bibitem{BGI}
Reuven Bar-Yehuda, Oded Goldreich and Alon Itai.
{\em On the time complexity of broadcast in radio networks: an exponential gap between determinism and randomization},
Journal of Computer and System Sciences 45 (1992), 104-126.

\bibitem{BGIE}
{\em Errata regarding "On the Time-Complexity of Broadcast in Radio
Networks: An Exponential Gap Between Determinism and Randomization"},
Dec. 2002, available from
$http://www.wisdom.weizmann.ac.il/~oded/p_bgi.html$

\bibitem{BV04}
Carlos Brito and Shailesh Vaya.
{\em Improved lower bound for deterministic broadcasting in radio networks},
Theoretical Computer Science, Preprint.

\bibitem{BP}
B. Bruschi and M. Del Pinto.
{\em Lower bounds for the broadcast problem in mobile radio networks},
Distributed Computing 10(1997), 129-135

\bibitem{CF}
I. Chlamtac and A. Farago.
{\em Making transmission schedule immune to topology changes in multi-hop packet radio networks},
IEEE/ACM Trans. on Networking 2(1994), 23-29

\bibitem{CGGPR02}
B. S. Chlebus, L. Gasieniec, A. Gibbons, A. Pelc and W. Rytter.
{\em Deterministic broadcasting in unknown radio networks},
In Distributed Computing 15 (2002), 27-28.

\bibitem{CGL02}
M. Christersson, L. Gasieniec, and A. Lingas.
{\em Gossiping with bounded size messages in ad-hoc radio networks},
In Proceedings of Twenty Nineth International Colloquium on Automata, Languages and Programming (ICALP'02), LNCS 2380, 377-389.

\bibitem{CGOR00}
Bogdan S. Chlebus, Leszek Gasieniec, Alan Gibbons, Anna Ostlin and John M. Robson.
{\em Deterministic radio broadcasting.}
In Proceedings 27th International Colloquium on Automata, Languages and Programming, ICALP'2000,LNCS 1853, 717-728.

\bibitem{CGR02}
Marek Chrobak, Leszek Gasieniec and Wojciech Rytter.
{\em Fast broadcasting and gossiping in radio networks},
In Proceedings 41st Annual IEEE Symposium on Foundations of Computer Science, FOCS'2000, 575-581.


\bibitem{CK},
B.S. Chlebus and Dariusz Kowalski.
{\em A better wake-up in radio networks},
In Proceedings of Annual Symposium on Principles of Distributed Computing, PODC 2004.


\bibitem{CMS}
Andrea E.F. Clementi, Angelo Monti and Riccardo Silvestri.
{\em Selective families, superimposed codes, and broadcasting on unknown radio networks},
In Proceedings 12th Annual ACM-SIAM Symposium on Discrete Algorithms (SODA'2001), 709-718.


\bibitem{CR}
Artur Czumaj and W. Rytter.
{\em Broadcasting Algorithms in Radio Networks with Unknown Topology},
In Proceedings of the 44th Annual IEEE Symposium on Foundations of Computer Science, FOCS 2003, Cambridge, MA.


\bibitem{CRM}
B.S. Chlebus and M. Rokicki.
{\em Asynchronous broadcasting in radio networks},
To appear in International Colloquium in Structural Information and Communication Complexity, (SIROCCO 2004), LNCS 3104, 57-68.

\bibitem{CRUZ}
R. Cruz and B. Hajek,
{\em A new upper bound to the throughput of a multi-access broadcast channel},
IEEE Transactions on Information Theory IT-28, 3 (May 1982), 402-405.


\bibitem{CW}
I. Chlamtac and O. Weinstein.
{\em The wave expansion approach to broadcasting in multihop radio networks},
IEEE Transactions on Communications 39(1991), 426-433.


\bibitem{EK}
M. Elkin and G. Kortsartz.
{\em Improved broadcast schedule for radio networks},
In Sixteenth Annual ACM-SIAM Symposium on Discrete Algorithms, SODA 2005.

\bibitem{FP08}
E. G. Fusco and A. Pelc.
{\em Acknowledged broadcasting in ad-hoc radio networks},
In Information Processing Letters, 109 (2008), 136-141.


\bibitem{G09}
L. Gasienec,
{\em On effecient gossiping in radio networks},
In Proceedings of Sixteenth International Colloquium on Structural Information and Communication Complexity, SIROCCO 2009, LNCS 5869, 2-14, Sirince, Turkey.


\bibitem{GL02}
L. Gasieniec and A. Lingas,
{\em On adaptive deterministic gossiping in ad hoc radio networks},
Information Processing Letters 83 (2002), 89-94.


\bibitem{M10}
Gianluca De Marco.
{\em Distributed Broadcast in Unknown Radio Networks},
SIAM Journal of Computing 39(6):2162-2175 (2010).

\bibitem{MP01}
Gianluca De Marco and Andrzej Pelc.
{\em Faster broadcasting in unknown radio networks},
Information Processing Letters 79 (2001), 53-56.


\bibitem{GM}
I. Gabour and Y. Mansour.
{\em Broadcast in radio networks},
In Proceedings 6th Annual ACM-SIAM Symposium on Discrete Algorithms, SODA 1996, 577-585.

\bibitem{GP}
L. Gasieniec and I. Potapov,
{\em Gossiping with unit size messages in known Radio Networks},
In Proceedings of Second International Conference on Theoretical Computer Science (TCS'02), Volume 223 IFIP Conference Proceedings, 193-205.

\bibitem{GPP}
L. Gasieniec, A. Pelc and D. Peleg.
{\em The wakeup problem in synchronous broadcast systems},
In SIAM Journal on Discrete Mathematics 14, 2001, 207-222.

\bibitem{GPP02}
L. Gasieniec, A. Pagourtzis, and I. Potapov.
{\em Deterministic communication in radio networks with large labels},
In Proceedings of Tenth Eurpoean Symposium on Algorithms, ESA 2002, LNCS 2461, 512-524.


\bibitem{GRX04}
L. Gasieniec, T. Radzik, and Q. Xin.
{\em Faster Deterministic Gossiping in Directed Ad Hoc Radio Networks},
In Scandian Workshop on Algorithmic Theory, SWAT 2004, 397-407, 2004.


\bibitem{H95}
F. K. Hwang.
{\em The time complexity of deterministic broadcast in radio networks},
Discrete Applied Mathematics 60(1995), 219-222.


\bibitem{JS}
T. Jurdzinski and G. Stachowiak,
{\em Probabilistic algorithms for the wake up problem in single-hop radio networks},
In Proceedings of 13th International Symposium on Algorithms and Computing, ISAAC 2002, LNCS 2518, 525-549.


\bibitem{KM}
Eyal Kushilevitz and Yishay Mansour,
{\em An $\Omega(D\log(N/D))$ lower bound for broadcast in radio networks},
SIAM Journal on Computing 27(1998), 702-712


\bibitem{KP04}
Dariusz R. Kowalski and Andrzej Pelc,
{\em Time of deterministic broadcasting in radio networks with local knowledge},
In SIAM Journal on Computing 33(4): 870-891(2004).

\bibitem{KP2}
Dariusz R. Kowalski and Andrzej Pelc,
{\em Faster Deterministic Broadcasting in Ad Hoc Radio Networks},
In Proceedings 43rd Symposium Theoretical Aspects of Computer Science, STACS 2003.

\bibitem{KP3}
Dariusz R. Kowalski and Andrzej Pelc,
{\em Time complexity of radio broadcasting: adaptiveness vs obliviousness and vs randomization vs determinism},
In Theoretical Computer Science 33(3): 355-371(2005)

\bibitem{KP4}
Dariusz R. Kowalski and Andrzej Pelc,
{\em Centralized deterministic broadcasting in undirected multihop radio networks},
In Proceedings of 7th International Workshop on Approximation algorithms for Combinatorial Optimization Problems, APPROX 2004, LNCS 3122,171-182.

\bibitem{KP5}
Dariusz R. Kowalski and Andrzej Pelc,
{\em Optimal Deterministic Broadcasting in Known Topology Radio Networks},
In Distributed Computing 19(3): 185-195 (2007).

\bibitem{KP6}
Dariusz R. Kowalski and Andrzej Pelc,
{\em Broadcasting in undirected ad hoc radio networks},
In Distributed Computing 18(1): 43-57(2005).

\bibitem{KS}
W.H.Kauz and R.R.C. Singleton,
{\em Nonrandom binary superimposed codes},
In IEEE Transactions on Information Theory 10(1964), 363-377.

\bibitem{LP}
D. Liu and M. Prabhakaran,
{\em On Randomized Broadcasting and Gossiping in Radio Networks},
In Proceedings of Eigth Annual International Conference on Computing and Combinatorics, COCOON'02, LNCS 2386, 340-349.

\bibitem{MX}
F. Manne and Q. Xin,
{\em Optimal Gossiping with Unit Size Messages in Known Topology Radio Networks},
In Combinatorial and Algorithmic Aspects of Networking, CAAN 2006, LNCS 2006, 125-135.

\bibitem{OCW03}
T. Okuwa, W. Chen, K. Wada,
{\em An optimal algorithm for acknowledged broadcasting in ad hoc radio networks},
In Proceedings of Second International Symposium on Parallel and Distributed Computing (2003), 178-184.

\bibitem{P}
D. Peleg,
{\em Deterministic radio broadcast with no topological knowledge},
Manuscript, 2000.

\bibitem{UCW07}
Jiro Uchida , Wei Chen, Koichi Wada,
{\em Acknowledged broadcasting and gossiping in ad hoc radio networks},
In Theoretical Computer Science, 377, 43-54 (2007).

\bibitem{Xu}
Xu, Y.,
{\em An $O(n^{3/2})$ deterministic algorithm for radio networks}
In Algorithmica 36(1), 93-96 (2003).
}
\end{thebibliography}
\end{document}